\documentclass[12pt]{article}
\usepackage{ amssymb }
\usepackage{graphicx}
\usepackage{float}
\usepackage{hyperref}
\graphicspath{ {figures/} }
\usepackage{array}
\begin{document}
\pagenumbering{gobble}

\Large
\begin{center}
 	$\phi$-Lagrangian, a new scalar mediator for light-by-light scattering process\\ 
	
	\hspace{10pt}
	
	\large
	 BELFKIR Mohamed$^1$,
	
	\hspace{10pt}
	
	\small
	$^1$\href{mailto:mohamed.belfkir@cern.ch}{mohamed.belfkir@cern.ch}\\ High Energy Physics and Astrophysics Laboratory LPHEA,\\ Faculty of Science SEMLALIA,  \\ Marrakesh, Morocco \\  
\end{center}

\hspace{10pt}

\normalsize
\textbf{Abstract}. {\em The $\phi$ scalar boson is a new mediator proposed to describe the direct light-by-light scattering observed recently in ultra-peripheral collisions (UPCs) with ATLAS detector where two photons interact directly to gives two photons in the final state. The proposed lagrangian present a description to the forbidden process. The description presented in this paper consist the interaction terms, the coupling to the Higgs boson to generate the mass for this new resonance and the simulation using proton proton in order to observe this process in other nominal LHC collisions, since the process is observed in heavy-ions collisions.   }
\tableofcontents
\thispagestyle{empty}
\listoffigures
\pagenumbering{arabic}
\section{Introduction}
\begin{figure}[ht]
	\centering
	\includegraphics[width=0.3\linewidth]{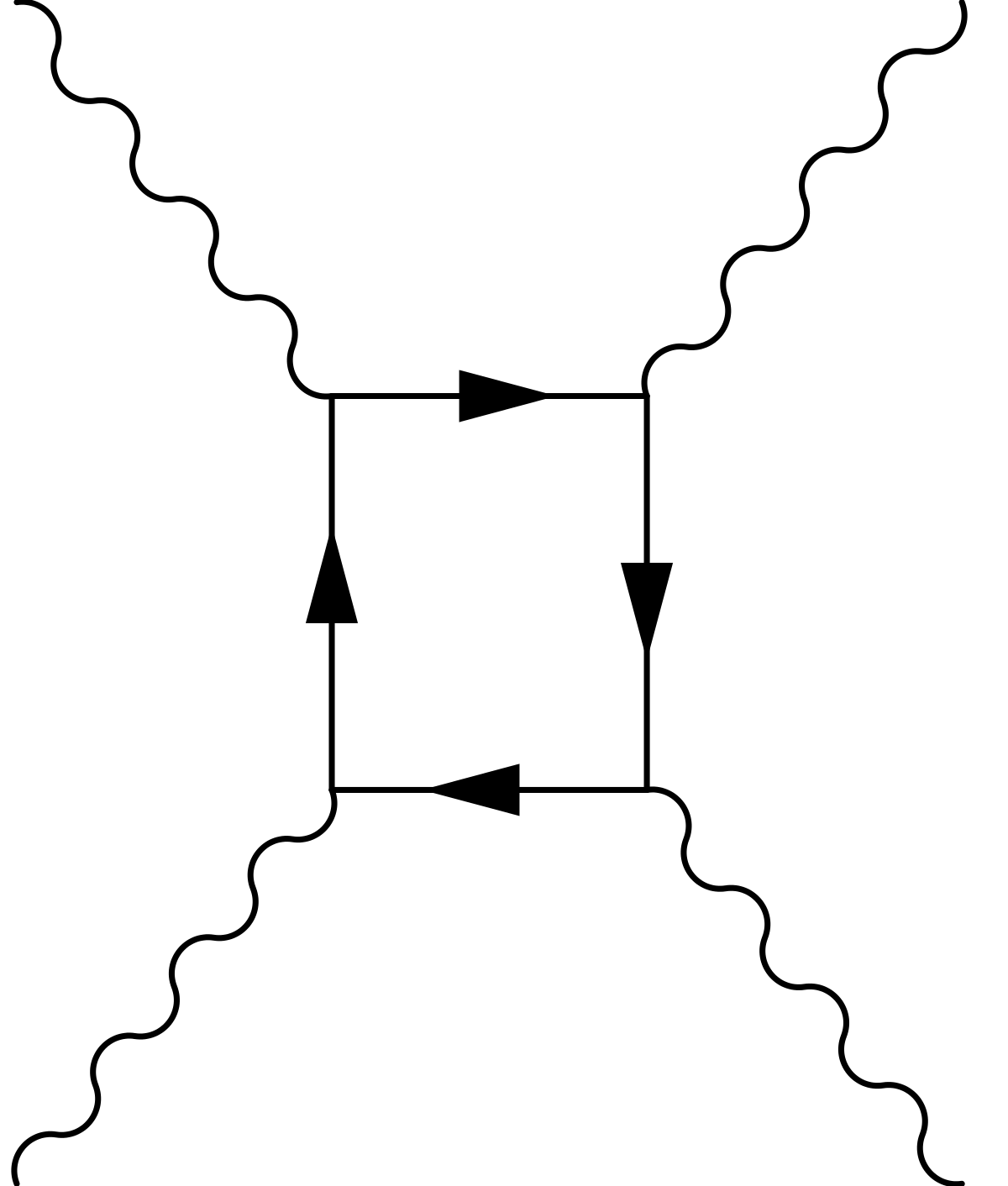}
	\caption{Elastic QED light-by-light scattering Feynman diagram}
	\label{fig:photon-photonscattering}
\end{figure}

Recently, a direct light-by-light scattering $(\gamma\gamma\rightarrow\gamma\gamma)$ phenomena is observed with ATLAS detector at LHC in heavy-ion collisions \cite{1,2}. Thirteen events of this highly suppressed  classical-electrodynamic process were observed with an expected background of $2.6 \pm 0.7$ events. In the Standard Model $(SM)$ of particle physics this process can be described with a box Feynman diagram with all charged particles in the box (figure \ref{fig:photon-photonscattering}). In order to consistently explain ATLAS results, a phenomenological  model is described in this paper. In Section 2, the effective Lagrangian associated to a new scalar boson mediator between the two photons vertex is described. This Lagrangian is derivated from the Axion-like Particles $(ALPs)$ Lagrangian presented in \cite{3} by changing the interaction term and generating the $ALPs$ mass using Higgs mechanism as discussed in Section 3. Next, in Section 4, a signal simulation using $MadGraph5\_aMC@NLO$ Monte Carlo generator \cite{4} is treated to describe the kinematic of this $\phi-boson$.\\ 

\section{The $\phi$ Lagrangian}
\begin{figure}[ht]
	\centering
	\includegraphics[width=0.45\linewidth]{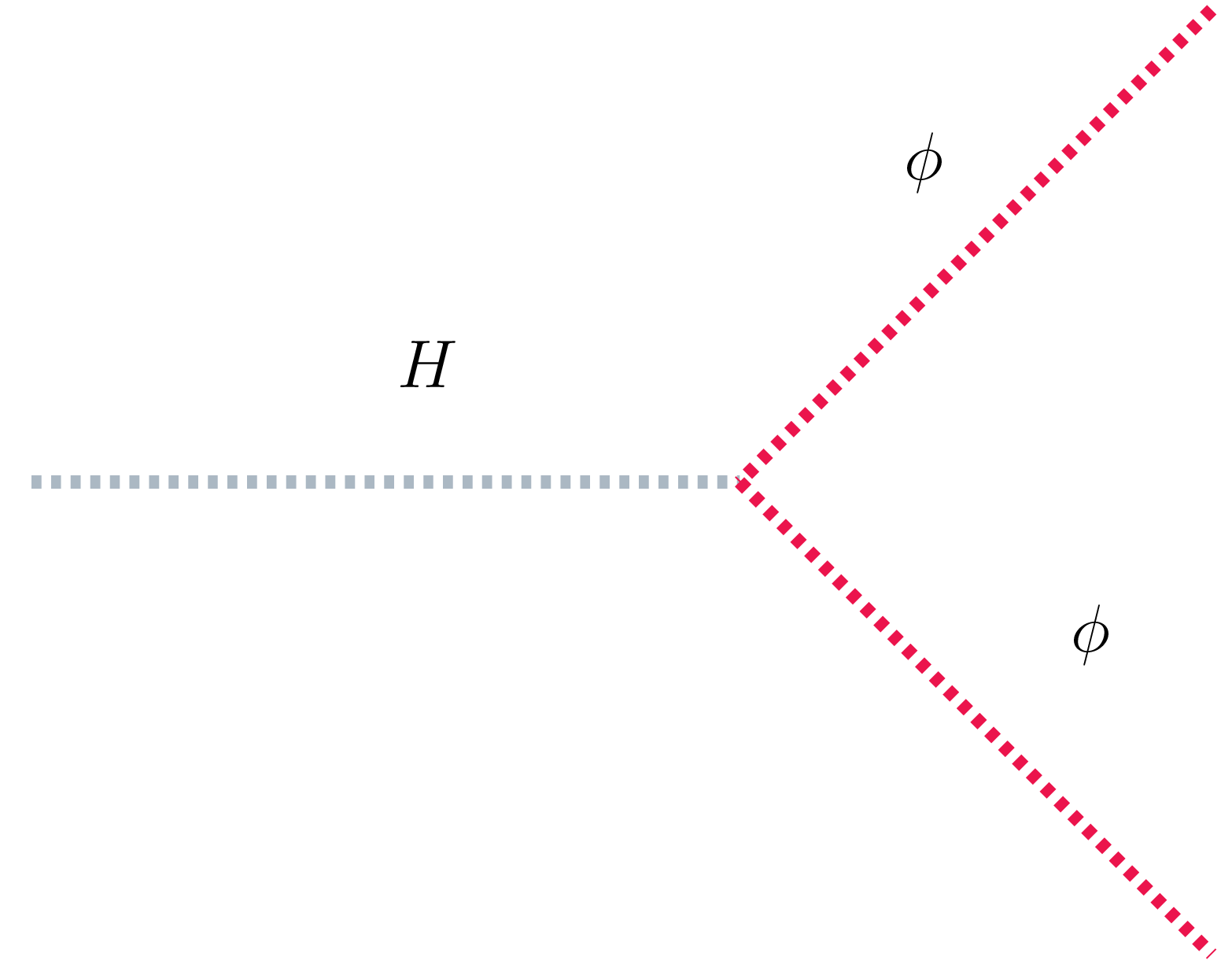}
	\caption{$\phi$-exchanging feynman diagram}
	\label{fig:diagram2}
\end{figure}

As abraded mentioned, the Lagrangian describing the light-by-light scattering is deriveted from \cite{3} by introducing a new interaction terms using directly the interaction between hyper-photon field $B^{\mu}$ before the symmetry breaking and the new scalar $\phi$. The $\phi$ exchange (figure \ref{fig:diagram2}) is described by the following lagrangian density: \\
\begin{equation}
	\mathcal{L} =\frac{1}{2}(\partial\phi)^2-\frac{1}{2}M_{\phi}^2\phi^2- \frac{\lambda}{2!}\phi B^{\mu}B_{\mu} - \frac{1}{4}F^{\mu\nu}F_{\mu\nu}
	\label{eq:eq1}
\end{equation}
where $F^{\mu\nu} $ is the electromagnetic field strength tensor calculated using the $B^{\mu}$. $\lambda$ and $M_{\phi}$ are the two free parameters in this model, the coupling constant and the $\phi$ mass respectively. Since the $\phi$ mass is generated through the Brout-Englert-Higgs mechanism. very simple Yukawa coupling to Higgs bosons, by exploiting the invisible Higgs width decay we set a upper limits on the $\phi$ mass (See Section 3). \\
Using eq\ref{eq:eq1}, the full width decay of $\phi$ to di-photon system is :
\begin{equation}
	\Gamma_{\phi\rightarrow\gamma\gamma} = \frac{\cos^4\theta_w\lambda^2}{2\pi M_{\phi}},
	\label{eq:eq2}
\end{equation}
where $\theta_w$ is the Weinberg angle.\\
The $\phi$ scalar can also decay very weakly into two off-shell $Z$ boson, this rare process can be useful to prove the existence of the $\phi$ boson. \\
\section{$M_\phi$ generation}
\begin{figure}[ht]
	\centering
	\includegraphics[width=0.3\linewidth]{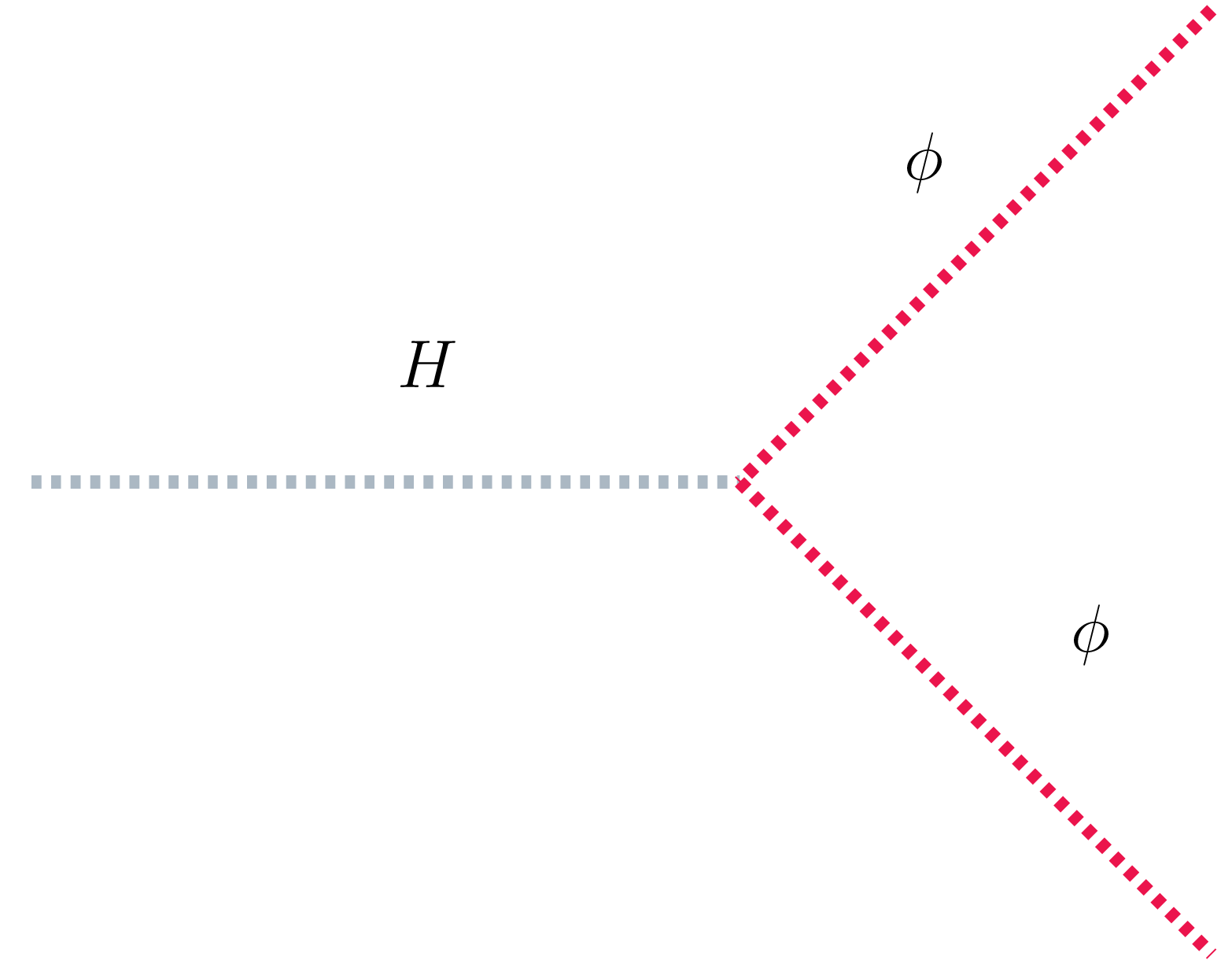}
	\caption{Higgs boson decay to $\phi$}
	\label{fig:diagram3}
\end{figure}

In order to make this model renormalizable the $\phi$ mass was generated using a very simple coupling to the  Higgs field as following :
\begin{equation}
		\mathcal{L}_{mass} = \frac{1}{2}M_{\phi}^2\phi^2 = \gamma\phi\Phi^+\phi\Phi,
		\label{eq:eq3}
\end{equation}
where $\Phi =\frac{1}{\sqrt{2}}(0 \ vev + h(x))^+$ is the Higgs field, $h(x)$ is the excitation of vacuum which correspond to the Higgs boson field, $vev$ is the vacuum expectation value of the Higgs potential ($\sim 246 \ GeV$), and $\gamma$ is dimensionless coupling constant between the Higgs boson and the $\phi$ boson. By developing (eq \ref{eq:eq3}) is possible to derive the mass terms and all interaction vertices between the Higgs and the $\phi$ boson:
\begin{equation}
	\mathcal{L}_{mass} = \frac{1}{2}\gamma v^2\phi^2 + \frac{1}{2}\gamma\phi^2h^2 + \gamma v \phi^2h
	\label{eq:eq4}
\end{equation}
Eq \ref{eq:eq4} shows that $M_\phi^2$ is now equal to $\gamma v^2$, and by using the Higgs boson decay (figure \ref{fig:diagram3})we can fix an upper limit on the $\gamma$ coupling, consequently an upper limit on the $\phi$ mass.

\subsection{Upper limit on $M_\phi$} 
The Higgs boson has a full width decay $\Gamma_H$ smaller than $1.7 GeV$ with Confidence Level $(CL) \ 95\%$ , and $\Gamma(H\rightarrow invisible) < 58\% \sim 0.9 GeV$ at 95\% CL \cite{5}. \\
Assume that the Higgs to invisible is just the decay of the Higgs to two $\phi$ bosons, that makes the $\Gamma(H\rightarrow\phi\phi) \sim 0.9 GeV$. and :
\begin{equation}
	\Gamma(H\rightarrow \phi\phi) = \frac{\gamma^2v^2}{4\pi M_H},
	\label{eq:eq5}
\end{equation}
where we neglect the $M_{\phi} < M_H$ in front of the mass of Higgs $(125 \ GeV)$. \\
using (Eq\ref{eq:eq5}), $\gamma$ will be $\sim 0.153$ and the $M_{\phi} \sim 96.17 \ GeV$. The limit on $M_{\phi}$ shows that in an ideal case, $\phi$ has a mass in the range $[0 - 96] \ GeV$.
In the case, the branching ratio of the Higgs boson to $\phi$ is just  $0.1\%$, the mass of $\phi$ is reduced to  $\sim  17 \ GeV$ and $\gamma \sim 4.8\times10^{-3}$.
As a result, a signature of di-Z boson scattering below the Higgs peak can be a powerful evidence of existence of new physics at low di-photon invariant mass.\\ 

\section{$\phi$-Signal Simulation}
To implement the full $\phi$ Lagrange density in $MadGraph5\_aMC@NLO$, $Feyn-Rules$ packages \cite{6} $Mathematica$ \cite{10} is used to create a UFO model, used to generate the $\phi$ exchanging signal. To describe reconstruction effect a additional smearing is applied using $Delphes$ \cite{7,8} with ATLAS card modified to allow reconstruction of low photon energy, the minimum $Pt_{\gamma}$ is set to be $4 \ GeV$. 

\begin{figure}[ht]
	\centering
	\includegraphics[width=0.7\linewidth]{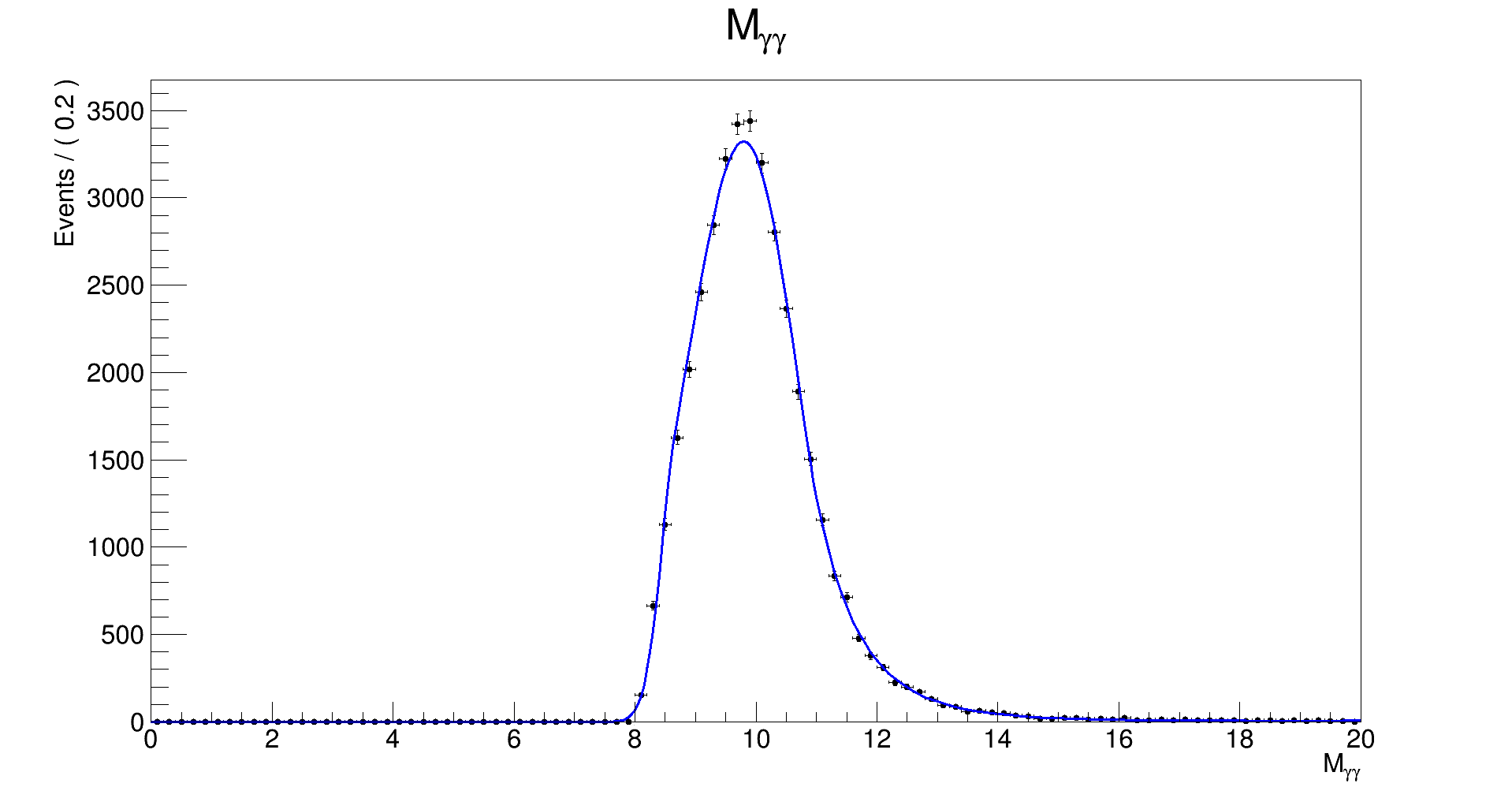}
	\caption{Di-photons invariant mass distribution fitted using Bukin function}
	\label{fig:myy}
\end{figure}

The di-photon invariant mass after simulation using the full $\phi$ Lagrangian, fitted using Bukin \cite{9} distribution, is shown in Fig \ref{fig:myy}. 



\section{Conclusion}
The presented phenomenological lagrangian describe very nicely the photon-photon scattering observed by the ATLAS collaboration in heavy-ions collisions, through a new scalar boson $\phi$, the forbidden process may be also observable in proton-proton (p-p) collisions, the cross section and the luminosity required to observe this resonance using p-p collisions will be studied later.

\end{document}